\documentclass[aps, prb, twocolumn, floatfix, amssymb]{revtex4}

\usepackage{graphicx}

\begin{document}

\title{Apparatus for high resolution microwave spectroscopy in strong magnetic fields}

\author{W. A. Huttema}
\author{B. Morgan}
\altaffiliation[Permanent address: Cavendish Laboratory, Madingley
Road, Cambridge, CB3 0HE, United Kingdom]{}
\author{P. J. Turner}
\author{W. N. Hardy}
\altaffiliation[Permanent address: ]{Dept. of Physics and Astronomy,
University of British Columbia, Vancouver, BC V6T 2A6, Canada}
\author{Xiaoqing Zhou}
\author{D. A. Bonn}
\altaffiliation[Permanent address: ]{Dept. of Physics and Astronomy,
University of British Columbia, Vancouver, BC V6T 2A6, Canada}
\author{Ruixing Liang}
\altaffiliation[Permanent address: ]{Dept. of Physics and Astronomy,
University of British Columbia, Vancouver, BC V6T 2A6, Canada}
\author{D. M. Broun}
\affiliation{Department of Physics, Simon Fraser University,
Burnaby, BC V5A 1S6, Canada}
\date{\today}

\begin{abstract}

We have developed a low temperature, high-resolution microwave surface impedance probe that is able to operate in high static magnetic fields. Surface impedance is measured by cavity perturbation of dielectric resonators, with sufficient sensitivity to resolve the microwave absorption of sub-mm-sized superconducting samples.  The resonators are constructed from high permittivity single-crystal rutile (TiO$_2$) and have quality factors in excess of $10^6$.  Resonators with such high performance have traditionally required the use of superconducting materials, making them incompatible with large magnetic fields and subject to problems associated with aging and power-dependent response.  Rutile resonators avoid these problems while retaining comparable sensitivity to surface impedance.  Our cylindrical rutile resonators have a hollow bore and are excited in TE$_{01(n-\delta)}$ modes, providing homogeneous microwave fields at the center of the resonator where the sample is positioned.
 Using a sapphire hot-finger technique, measurements can be made at  sample temperatures in the range 1.1~K to 200~K, while the probe itself remains immersed in a liquid helium bath at 4.2~K.  The novel apparatus described in this article is an extremely robust and versatile system for microwave spectroscopy, integrating several important features into a single system.  These include: operation at high magnetic fields;  multiple measurement frequencies between 2.64~GHz and 14.0~GHz in a single resonator; excellent frequency stability, with typical drifts $<$ 1~Hz per hour; the ability to withdraw the sample from the resonator for background calibration; and a small pot of liquid helium separate from the external bath that provides a sample base temperature of 1.1~K.  Without modification, this system can be employed for dielectric spectroscopy, electron spin resonance and other microwave spectroscopies.
\end{abstract}

\maketitle

\section{Introduction}

Microwave spectroscopy is a powerful probe of the long-wavelength
electromagnetic response of solids, coupling to conduction electrons
as well as electric and magnetic excitations.  Although we are
chiefly concerned with the low energy electrodynamics of
superconductors,  the apparatus we describe in this article can be
used without modification for dielectric spectroscopy and electron
spin resonance, at temperatures from 1.1~K to 200~K.   This system was developed for microwave
spectroscopy in strong magnetic fields, but is sensitive enough to
allow high-resolution measurements on sub-mm-sized samples of materials with extremely low microwave absorption, such as superconductors and low-loss dielectrics.

Microwave spectroscopy is particularly useful for superconductors,
in which the d.c.~resistivity is shorted to zero by supercurrents.
At microwave frequencies the conductivity of a superconductor
remains finite but develops a signficant out-of-phase
component due to the purely reactive response of the
super-electrons.  Microwave experiments on superconductors fall into
two categories, depending on the strength of the applied static magnetic
fields.  In the Meissner state, magnetic fields are excluded from the
bulk of the sample.  In this regime the utility of microwave
spectroscopy lies in the dual character of the complex conductivity,
$\sigma = \sigma_1 -  i \sigma_2$.  The real part $\sigma_1$
contains details of the scattering dynamics of thermally excited
quasiparticles, or `normal' electrons. The imaginary part $\sigma_2$ acts as a
thermodynamic probe of the electronic excitation spectrum, its value at low frequencies being proportional to  the temperature-dependent superfluid density.  In a
type-II superconductor strong magnetic fields drive a transition to
the vortex state, in which the field penetrates the superconductor
as quantized lines of magnetic flux.  The microwave response of the
vortex state is also of great interest, for both fundamental and
practical reasons.  On the fundamental side, microwave spectroscopy
provides a clean method of measuring electrical transport properties of the quasi-normal excitations associated with the vortex cores.\cite{morgan05,coffey91}  This is especially important
for the cuprate superconductors, in which upper critical fields are
so large that the normal state remains inaccessible in even the
highest laboratory fields.  On the practical side, the same
experiments  measure the pinning forces experienced by flux lines,
and the effect of temperature on depinning them.

For microwave spectroscopy of conducting materials the experimentally
accessible quantity is the surface impedance $Z_s =
R_s + i X_s$, the ratio of the transverse components of electric and magnetic field at the surface
of the sample.  In metals and superconductors where local electrodynamics hold, $Z_s$ is directly related to the conductivity, $\sigma = i \omega \mu_0/Z_s^2$. In principle, the surface impedance can  be obtained from an
optical-type reflectivity experiment.  In practice this approach has
limited success, because in good metals and superconductors the reflectivity
at microwave frequencies is indistinguishable from unity.
In microwave work, reflectivity measurements are further complicated by diffraction
effects, since the wavelength becomes comparable
to the size of a typical sample.  The low-frequency approach, in
which  leads are directly attached to the sample, is also
problematic at microwave frequencies, because the impedance of the sample is
typically only a fraction of an ohm.  This would be comparable to the contact resistance and be difficult to match to the characteristic impedance of a microwave transmission line.

A widely adopted solution in the microwave range is cavity
perturbation, which circumvents the limitations of both the optical
and low-frequency approaches.  In cavity perturbation the sample is
treated as part  of the inner surface of a  high quality factor
(high-$Q$) resonator, often a superconducting cavity.  As we will show below, the surface impedance
can be directly inferred from a measurement of the frequency and line-width of one of
the resonator modes.   The resonator amplifies the interaction of
electromagnetic radiation with the sample by the $Q$ factor, through
repeated reflection of the fields.  This multiplies up the signal
and provides effective impedance matching to transmission lines.   As a result, the resolution of the experiment improves with the $Q$ of the resonant mode, allowing very low loss samples to be measured.  The
resonator also creates fields that are spatially well defined,
simplifying geometric effects that arise in the diffraction limit.
In the case of electrically anisotropic materials, experiments are
simplified by judicious choice of the resonant mode, as it is
usually possible to align the microwave field polarization with one
of the principle axes of a single-crystal sample.

Current research in superconductivity often has to deal with materials
having complex chemistry, such as the high-$T_c$ cuprates, heavy
fermion compounds and organic superconductors.  In most cases the
best quality samples are small single-crystal specimens, often less than a
millimeter in size.  Cavity perturbation measurements on such
samples are difficult because they have low microwave absorption, and because the sample surface comprises a
very small fraction of the area of the resonator walls.  The challenge of
accurately separating  temperature-dependent contributions due to the sample from the resonator background was solved by the introduction of a sapphire hot-finger
technique,\cite{sridhar88} an innovation that has since been widely
adopted.   In this technique, the sample is positioned
inside the resonator on an independently heated sapphire plate that is thermally isolated
from the resonator.  (See Fig.~\ref{fig:MMPfull}.) Sapphire is a good thermal conductor and has low microwave absorption, allowing the sample temperature to be
monitored and controlled from outside the resonator, without the thermometry circuits
interfering with the operation of the resonator.  The resonator temperature is held
fixed during the experiment, so that all changes in
the resonator frequency and line-width can be attributed to the temperature dependence of the sample's surface impedance.

For measurements at low temperatures in zero static magnetic field,
resonators for cavity perturbation are usually constructed using
superconducting materials, either from bulk material such as niobium
metal,\cite{sridhar88} or from copper with a superconducting
coating, often a PbSn alloy.\cite{dietl89,bonn91}  At high
microwave frequencies resonators are usually cylindrical cavities
operating in the TE$_{011}$ mode, with the sample positioned on the
cylinder axis along an $E$-field node, at an $H$-field antinode.
Below 10~GHz TE$_{011}$ resonators become quite large, and the
accompanying decrease in filling factor makes them unsuitable for
measurements on small single crystals.  Better filling factors are
obtained using more compact resonators, with split-ring and
loop--gap geometries being common choices.\cite{bonn91}  Superconducting
resonators achieve quality factors ranging from $10^6$ for
split rings\cite{bonn91} to over $10^{10}$ in the best cylindrical
cavities,\cite{turneaure68} which makes them excellent at measuring samples
with extremely low microwave absorption.  Problems with
superconducting resonators include the need for complex chemical
treatments when preparing the cavity
surfaces,\cite{turneaure68,dietl89,bonn91} and the tendency for
joints in the resonator to act as weak superconducting links with 
highly nonlinear microwave response and degradation over time.  The latter effect can lead to
a strong dependence of the resonator $Q$ on microwave field
strength, and often limits the operation of the resonator to low
microwave powers.  The quality of these resonators also tends to
decrease over time due to surface oxidation of the
superconducting material.  However, for the most sensitive cavity
perturbation measurements, high quality factors are essential and,
until now, there has been no alternative to superconducting
resonators.

For measurements in an applied magnetic field
superconducting materials are no longer an option as they are
either driven normal by the field or enter a highly dissipative
state of flux flow.  One solution is to construct resonators from
high conductivity, oxygen-free copper but, even at low temperatures, $Q$ factors are limited to
the order of  1$\times10^4$ to  5$\times10^4$, depending on frequency and geometry.\cite{bonn91,silva98}    A
different approach, giving substantially better results, is to use
shielded dielectric resonators.  In this case the dielectric
confines almost all the microwave magnetic flux, and screening currents
in the surrounding metal are greatly reduced.  In previous work
Gough and coworkers made hollow sapphire resonators for cavity
perturbation that operate in the range 10 to 20~GHz, and reported $Q$
factors of 2$\times10^5$ to  3$\times10^5$ at low temperature.\cite{gough01}  In the present work we
show that better filling factors and higher $Q$'s can be obtained
using high permittivity rutile (TiO$_2$) dielectric.  The microwave properties of rutile are extraordinary.\cite{luiten98}   At low temperatures, in the basal plane, it has a relative dielectric constant of approximately 120
and a loss tangent as low as $3 \times 10^{-8}$.  Here we report results on two copper-enclosed
rutile resonators operating in the frequency range  2.64 to 14.0~GHz.  These have modes with quality
factors between 1.1$\times10^6$ to  6.0$\times10^6$ and, in the case of the smaller resonator, a filling factor better than that of a 40~GHz cavity resonator.  This results in
surface-impedance resolution comparable to that of the best
superconducting cavity resonators, with the added capability of operation in high magnetic fields.  In
addition, copper-shielded dielectric resonators avoid all problems
associated with aging and power dependence in superconducting
resonators.  Superior frequency stability, with typical drifts less
than 1~Hz per hour, is also obtained, without the need to regulate
the temperature of the 4.2~K helium bath.  This is due to the
combination of  high mechanical stability of the resonator and the
weak temperature dependence of the resonator frequency in this range.

\section{Low Temperature Apparatus}

Our design for the high field cavity-perturbation apparatus is constrained by several requirements.  We need a mechanically rigid mount for the dielectric resonator, that puts the resonator in good thermal contact with the helium bath.  The resonator must be surrounded by a metallic enclosure, to prevent radiation losses, but the walls of the enclosure have to be kept well away from the resonator to minimize conduction losses.  The enclosure must contain two coupling ports, so that microwaves can be coupled through the resonator in transmission, and allow access for the sample, mounted on a sapphire hot finger.  As the hot finger and sample will be heated to temperatures in excess of 200~K, the space inside the enclosure has to be kept under high vacuum. Working in high magnetic fields places further constraints on the design.  The apparatus described here fits inside a standard 2$''$ diameter magnet bore, tightly restricting the space available.  The magnet used in this experiment, an 8~T NbTi superconducting solenoid from American Magnetics, has vapor-cooled leads, preventing the helium bath from being pumped to 1.2~K.   Instead, these temperatures are reached in our apparatus with a miniature 1~K pot operating in one-shot mode to cool the sample thermal stage.  In addition, the design includes a feature that is very desirable in cavity perturbation --- a movable sample stage that allows the sample to be completely withdrawn from the resonator.  This is done in our system by moving the whole assembly of 1~K pot and sample thermal stage.  As a result, the background microwave absorption of the resonator can be measured directly, allowing the {\em absolute} microwave absorption of the sample to be determined {\em in situ}.

\begin{figure}[t!]
\includegraphics*[width=88mm]{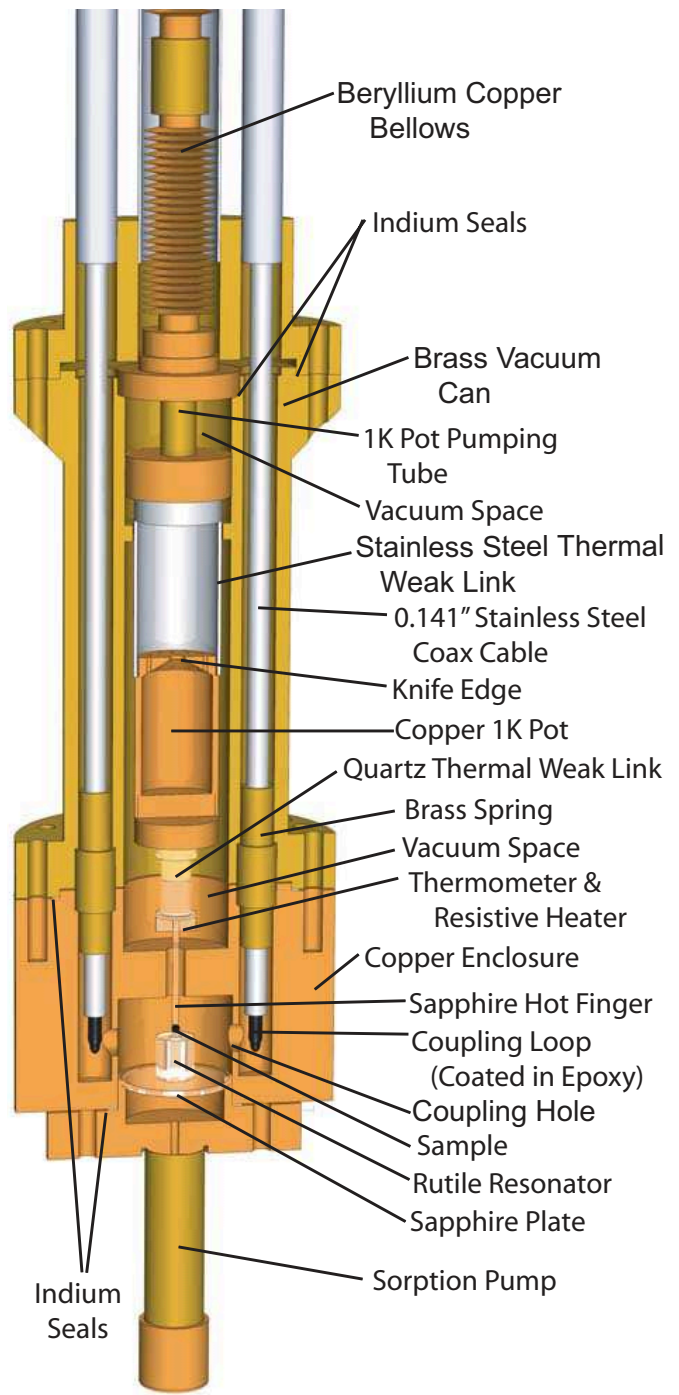}
\caption{\label{fig:MMPfull} (Color online) A cut-away schematic of the low temperature end of the cavity perturbation apparatus.   The 5.49~GHz rutile resonator is shown inside the copper enclosure, the outside of which is in direct contact with the helium bath at 4.2~K.  A movable sample thermal stage is one of the novel features of the design, consisting of an independently heated sapphire hot finger mounted below a movable 1~K refrigerator.  A key component of the motion stage is the flexible vacuum seal provided by BeCu bellows housed inside the 1~K~pot pumping line, separating it from the vacuum space.}
\end{figure}

The low temperature end of the apparatus is shown in
Fig.~\ref{fig:MMPfull}.  The cylindrical dielectric resonator, with an access
hole bored down the cylinder axis, is mounted with GE varnish onto a
sapphire plate that suspends it inside the copper enclosure.  The strength and high thermal conductivity of sapphire ensure rigid mounting and  good thermal contact
between the resonator and helium bath.  A pair of 0.141$''$ stainless-steel coaxial lines run the length of
the insert and terminate in loops near the cavity.  Microwaves are
coupled to the resonator through a pair of coupling ports in the
sides of the enclosure. The amount of coupling can be varied
by adjusting the vertical position of the coaxial lines, which enter the vacuum space through room-temperature sliding o-ring seals.  In our experiment the maximum coupling is limited by the size of the coupling holes, typically to about 2\% of critical coupling.  We usually optimize the position and orientation of the coupling loops to operate at this limit, as this improves signal strength and frequency stability.  It was found in early experiments that the Teflon dielectric in the coaxial cables allowed a slow gas leak from room temperature, which, if left unchecked, degraded the operation of the thermal stage.  These leaks were eliminated by coating the ends of the coaxial lines, including the coupling loops, with Stycast 2850 black epoxy.  A small charcoal sorption pump attached to the base of the copper enclosure further ensures high cryogenic vacuum for the duration of the experiment.

The temperature of the sample is controlled using a sapphire hot finger connected to  a small, isothermal sapphire platform that is weakly thermally connected to the base of the 1~K pot by a thin-walled quartz tube.   Sample temperature is measured with a Cernox thermometer,\cite{cernox} and is regulated between 1.1~K and 200~K by a Conductus/Neocera LTC-20 temperature controller.  The accuracy of the temperature control is $10^{-4}$~K below 10~K and $2\times10^{-3}$~K at the highest temperatures.  Wiring to the thermal platform (not shown in Fig.~\ref{fig:MMPfull}) passes through the 1~K pot for optimal thermal anchoring,  and there is no discernible heat leak from room temperature into the thermal stage.  Electrical feedthroughs between the helium and vacuum spaces are made with brass wires, sealed with Stycast 2850 epoxy as they pass through the copper base of the 1~K pot.  The entire thermal stage, including the 1K pot, is movable, allowing the sample to be completely withdrawn from the resonator for background calibrations. To permit motion while maintaining a hermetic seal between the 1~K pot pumping tube and the vacuum space, we use flexible BeCu bellows.\cite{bellows}  The motion is controlled by a stainless steel rod that runs up the inside of the 1~K pot pumping line and passes through a sliding o-ring seal at room temperature.  A ledge in the 1~K pot housing provides positive location for the sample stage when it is in the fully engaged position.

The 1~K pot is constructed in two parts to avoid thermal losses associated with superfluid film creep.  The lower section is made from copper and has a volume of 1.9~ml.  The upper section is a tube made from 0.001$''$ stainless-steel shim stock that acts as weak thermal link to 4.2~K.  The two sections are connected by a small, circular orifice, 0.5~mm in diameter with knife-like edges, that interrupts superfluid film flow.  The 1~K pot is operated in one-shot mode, with a charge of helium provided by connecting the 1~K pot to a supply of high-purity helium gas that liquifies to fill the lower section of the pot.  We find that the 1~K pot has a better base temperature and longer hold time if we assiduously keep the inside surfaces free of contamination.   An important part of this process is to pass the helium gas through a liquid nitrogren cold trap before it enters the apparatus in order to scrub it clean of impurity gases.  With these precautions,  boil-off from the 1~K pot is low enough that it can be pumped with a small turbo-molecular pump down to pressures below $10^{-3}$~mbar.  For operation below 4.2~K, a typical hold-time for the 1~K pot is 10~hours, with a base temperature below 1.1~K.  When operating above 4.2~K we keep a small overpressure of helium gas in the 1~K pot pumping line to ensure a reproducible thermal environment in the 1~K pot.

Sample interchange is performed with the low temperature insert warmed to room temperature.  The copper resonator enclosure is unbolted from the brass vacuum can at the second lowest indium seal and removed.  This gives direct access to the sapphire hot finger and thermal stage.  Several precautions are taken to protect the delicate hot-finger assembly during this procedure.  Before working near the sapphire rod, an aluminum mounting jig is brought into position and bolted into place on the indium flange at the bottom of the brass can.  Two $\frac{1}{8}''$ diameter brass guide pins, which protrude $2''$ from the base of the brass can, provide alignment for both the mounting jig and the resonator enclosure.  These work by mating with corresponding clearance holes drilled vertically through the jig and enclosure, guaranteeing alignment to better than 0.1~mm.  The sample to be measured is attached to the sapphire hot finger with a small amount of high vacuum grease,\cite{grease} which provides good thermal contact and has negligible microwave absorption.  The sample is mounted so that it sits at the center of the resonator with the 1~K pot stage in the fully engaged position.  This position is a local maximum in the on-axis microwave magnetic field intensity, as described in Section~\ref{sec:resonator}.  Siting the sample in this way enhances field uniformity and minimizes the contribution of sapphire thermal expansion to resonator frequency shift. 

\section{Cavity Perturbation}
\label{sec:cavitypert}

Cavity perturbation is a method for relating changes in the free response of a high $Q$ resonator to changes in the surface impedance $Z_s$, the permittivity $\epsilon$ and the permeability $\mu$ of small samples placed inside it. As shown in the Appendix, the main cavity perturbation result is:
\begin{eqnarray}
  \lefteqn{\Delta f_0 + i\Delta f_B/2\approx\bigg\{\frac{i}{2 \pi}\int_S\Delta Z_s\mathbf{H_1\cdot H_2}\mathrm{d}S}\nonumber
  \\&& - f_0\int_V[\Delta\mu
    \mathbf{H}_1\cdot\mathbf{H}_2+\Delta\epsilon'\mathbf{E}_1\cdot\mathbf{E}_2]
    \mathrm{d}V \bigg\}\bigg/4 U.\hspace{0.5cm}
  \label{eqn:cavitypert2}
\end{eqnarray}
Here $\Delta f_0$ and $\Delta f_B$ are the perturbative shifts in resonant frequency and bandwidth, respectively, and $U$ is the electromagnetic energy stored in the resonator.  $\textbf{E}$ and $\textbf{H}$ are complex phasor amplitudes of the electromagnetic fields, and the subscripts 1 and 2 denote the configurations before and after the perturbation.  The prime on the permittivity indicates that conduction currents have been folded into a redefinition of the dielectric response, for mathematical convenience.

 In our experiment the perturbation is caused by the presence of the sample and sapphire hot-finger, and is due to either a shift in sample temperature or a change in position.  For  surface impedance measurements on conductors the sample is inserted into the resonator along a node in the electric field. As a result, once the sample is fixed in place, the volume integral term in Eq.~\ref{eqn:cavitypert2} can usually be ignored, and the cavity perturbation formula reduces to
\begin{equation}
R_s(T) + i \Delta X_s(T) \approx \Gamma \big[\Delta f_B(T)/2 - i \Delta f_0(T)\big].
\end{equation}
 Here $\Gamma  = \int_S\textbf{H}_1\cdot\textbf{H}_2\mathrm{d}S/8\pi U$ is the resonator constant, which can typically be determined to an accuracy of 2.5\% using replica samples of known surface impedance.  
 $\Delta f_B(T)$ is the change in bandwidth on inserting the sample at temperature $T$ into the empty resonator, and $\Delta f_0(T)$ is the shift in resonant frequency on warming the sample from base temperature to $T$. It is never possible to determine the absolute reactance by inserting the sample, because in doing so the magnetic permeability term in Eq.~\ref{eqn:cavitypert2} gives a large frequency shift that dwarfs the contribution from surface reactance.  

\section{Dielectric Resonators}
\label{sec:resonator}

\begin{figure}[t]
\includegraphics*[width=65mm]{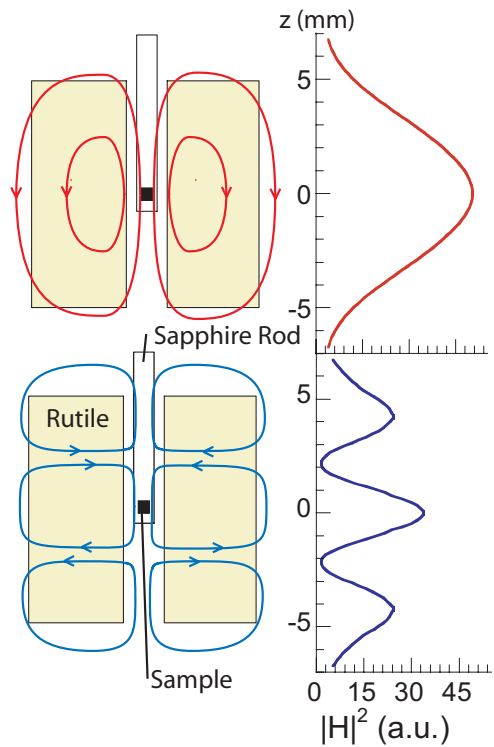}
\caption{\label{fig:MMPRutile} (Color online) An illustration of the magnetic field structure of the TE${}_{01(n-\delta)}$ modes, for $n = 1$ and $n = 3$.  The subscript $n-\delta$ specifies the number of half wavelengths of the TE$_{01}$ waveguide mode along the $z$ direction.  The TE${}_{01(1-\delta)}$ mode is sketched in the upper part of the diagram, with the TE${}_{01(3-\delta)}$ mode below it.  Alongside each illustration is a plot of the experimentally determined magnetic field intensity $|H|^2$, for the larger rutile resonator.  The $|H|^2$ data  were obtained from the change in bandwidth on inserting the sample, using the bead-pull procedure described in the text. }
\end{figure}

The resonators for this experiment are fabricated from oriented single crystals of rutile-phase TiO$_2$.  Rutile was chosen as the dielectric material for its high dielectric constant, $\epsilon_r \approx 120$, and its low loss tangent, $\tan \delta \approx 3 \times 10^{-8}$.  For a given frequency, the linear size of a dielectric resonator shrinks as $1/\sqrt{\epsilon_r}$, with a corresponding increase in  filling factor that increases as $\epsilon_r^{3/2}$.     For high $\epsilon_r$ material,  displacement currents are very  efficient at confining the microwave fields, reducing the magnitude of lossy conduction currents in the walls of the resonator enclosure.  The low loss tangent results in quality factors well in excess of a million.  

Our first resonator is a single-crystal rutile cylinder of radius 5.1~mm and height 5.9~mm, with a 1.5~mm diameter hole bored though the resonator along the cylinder axis.  The crystal c-axis is aligned with the cylinder axis to better than one degree, which is important because rutile is strongly birefringent. The raw single-crystal material for this resonator was supplied by eSCeTe,\cite{escete} and was ground and polished to shape by Microlap Technologies.\cite{microlap}  The fundamental frequency of this resonator is 5.49~GHz, and it has a $Q$ of $1.1 \times 10^6$ at 4.2~K.  Our second resonator is larger, with \mbox{diameter = height = 10~mm}, and a 3~mm diameter bore along the cylinder axis. Its fundamental resonance is at 2.64~GHz, with a $Q$ of $1.7 \times 10^6$ at 4.2~K.   Both the single-crystal growth and the resonator fabrication were carried out by ELAN.\cite{elan}

  The resonator modes used in our experiments are transverse electric modes with cylindrical symmetry, denoted TE${}_{01(n-\delta)}$ modes.  These are desirable for several reasons.  The electric field has a line node along the cylinder axis, minimizing the interaction of the sample and sapphire hot finger with electric fields. For the odd-order modes, the on-axis microwave magnetic field has maximum intensity at the center of the resonator.  The TE${}_{01(1-\delta)}$ and  TE${}_{01(3-\delta)}$ modes are sketched in Fig.~\ref{fig:MMPRutile}, showing the sample located at an antinode of the microwave magnetic field.  
  
 Modes are identified using the movable sample stage to carry out a so-called ``bead-pull'' experiment.  Although this only probes the fields along the resonator axis, it provides sufficient information to locate the TE$_{01(n-\delta)}$ modes.  To see how this works, we break Eq.~\ref{eqn:cavitypert2} into separate expressions for the shift in resonant frequency and bandwidth.  Keeping only the dominant contributions, we ignore the small dielectric loss of the sapphire hot-finger and assume that magnetic field is completely excluded from the interior of the conducting sample, making its effective permeability $-\mu_0$.  We then have
\begin{eqnarray}
\Delta f_0 & \approx & \frac{f_0}{4 U} \bigg\{\int_{\rm sample}\!\!\!\!\!\!\!\!\!\!\!\!\!\! \mu_0 \mathbf{H}_1\cdot\mathbf{H}_2 \textrm{d}V \!- \!\int_{\rm sapphire} \!\!\!\!\!\!\!\!\!\!\!\!\!\!\!\!\!\epsilon_r \epsilon_0 \mathbf{E}_1\cdot\mathbf{E}_2 \textrm{d}V \bigg\}\\
\Delta f_B & \approx & \frac{1}{4 \pi U} \int_{\rm sample}\!\!\!\!\!\!\!\!\!\!\!\!\! R_s \mathbf{H}_1\cdot\mathbf{H}_2 \textrm{d}S.
\end{eqnarray}
\begin{figure}[ht]
\includegraphics*[width=80mm]{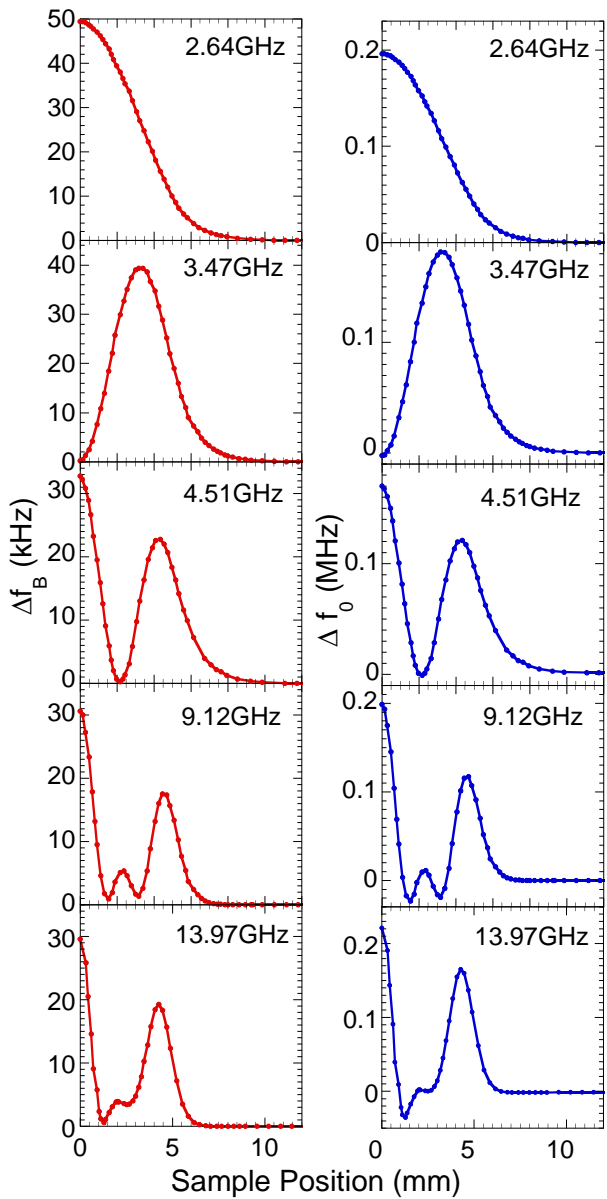}
\caption{\label{fig:f0andfbvspos} (Color online) Bead-pull results for a number of useful modes of the larger rutile resonator.  $\Delta f_B$ is the change in resonant bandwidth on inserting a small YBa$_2$Cu$_3$O$_{6.333}$ sample, in the normal state at 30~K, into the resonator.  $\Delta f_0$ is the shift in resonant frequency with respect to the empty resonator.  As described in the text, $\Delta f_B(z)$ gives a map of the magnetic field intensity along the axis.  $\Delta f_0(z)$ provides similar information, but contains a term due to the interaction of the sapphire hot finger with the electric fields of the resonator.  This term should be small for modes with an electric node along the cylinder axis.}
\end{figure}
Here the changes in $f_0$ and $f_B$ are with respect to the response of the empty resonator. The presence of a conducting sample reduces the effective volume of the resonator and increases its frequency.  The sapphire hot-finger has the opposite effect and, due to its large volume in comparison to the sample, gives a large negative frequency shift for any mode  not having an electric field node along the cylinder axis.  Most candidate modes can immediately be ruled out using this criterion alone.    Additional information is provided by the bandwidth, which for a conducting sample gives a clean measure of the magnetic field intensity.  Representative data is presented in Fig.~\ref{fig:f0andfbvspos} for the larger resonator, showing results for modes at 2.64, 3.47, 4.51, 9.12 and 13.97~GHz.  The empty resonator $Q$s of these modes are 1.73, 5.02, 6.0, 1.95 and 1.47~million respectively at 4.2~K.   Plots of $|H|^2$ vs. position obtained from the $\Delta f_B$ data are shown alongside the mode sketches in Fig.~\ref{fig:MMPRutile}.  A final confirmation of the field polarization comes from the use of a crystal of high-$T_c$ superconductor as the test sample.  The strong anisotropy of its electrodynamic response allows us to verify that the microwave magnetic field at the center of the resonator is indeed parallel to the cylinder axis.

\section{Microwave measurement circuit}

In our experiment we probe the response of the resonator modes in transmission, using an Agilent 8722ES vector network analyzer (VNA).  The VNA is set up to detect  the scattering parameter $\tilde{S}_{21}(f)$, the complex transmission amplitude through the microwave circuit.  For the free decay of a high $Q$ resonance, the dominant contribution to $\tilde{S}_{21}(f)$ is a simple pole at the complex decay frequency,
\begin{equation}
\tilde{S}_{21}(f) \sim \frac{1}{f - (f_0 + i f_B/2)}.
\end{equation}
A direct, non-resonant transmission amplitude $\tilde{D}_{21}$ acts in parallel with the resonant coupling.  The model used to fit to the measured scattering amplitude is
\begin{equation}
\tilde{S}_{21}(f) = \frac{\tilde{S}^0_{21}(f)}{1 - 2 i (f - f_0)/f_B} + \tilde{D}_{21},
\label{eqn:curvefit}
\end{equation}
where $\tilde{S}^0_{21}$ is the transmission amplitude on resonance.  Tildes denote complex quantities.

A LabVIEW routine processes the scattering parameter data from the VNA, fitting to them using a complex-valued Levenburg--Marquardt routine based on code from {\em Numerical Recipes in C}.\cite{numericalrecipes} This is an improvement over routines that fit only to the {\em amplitude} of the resonant response --- retaining the phase signal doubles the amount of information in each trace and makes direct coupling easy to account for.  For resonances with $Q$s of the order of $10^6$, the fitting routine resolves  $f_0$ and $f_B$ to better than 1~Hz.  Even at low $Q$ and in the presence of substantial direct coupling, Eq.~\ref{eqn:curvefit} models the data extremely well, only breaking down for samples with nonlinear absorption such as superconductors with weak links.

\begin{figure}[t]
\includegraphics*[width=\columnwidth]{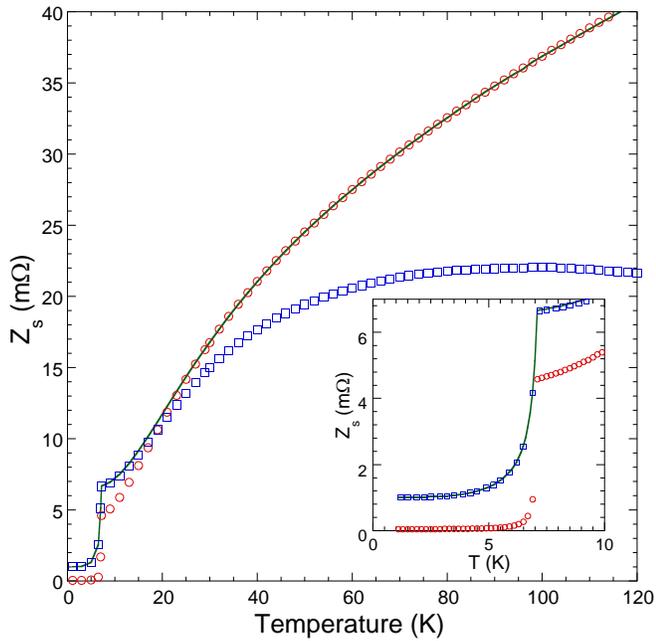}
\caption{\label{fig:PbRxXsvsT} (Color online)  5.49~GHz surface impedance data taken on a 350~$\mu$m diameter sphere of PbSn alloy.  Open circles are the surface resistance,
open squares are the measured surface reactance signal.  A thermal expansion correction is applied to the reactance data to obtain the solid line, as explained in the text. Once corrected, the temperature dependence of $R_s$ and $X_s$ track accurately, allowing the {\em absolute} reactance to be obtained by matching $R_s$ and $X_s$ in the normal state.  Inset: a close-up view of the surface impedance in the superconducting state, showing the weak temperature dependence expected at low temperature.}
\end{figure}

\begin{figure}[t]
\includegraphics*[width=\columnwidth]{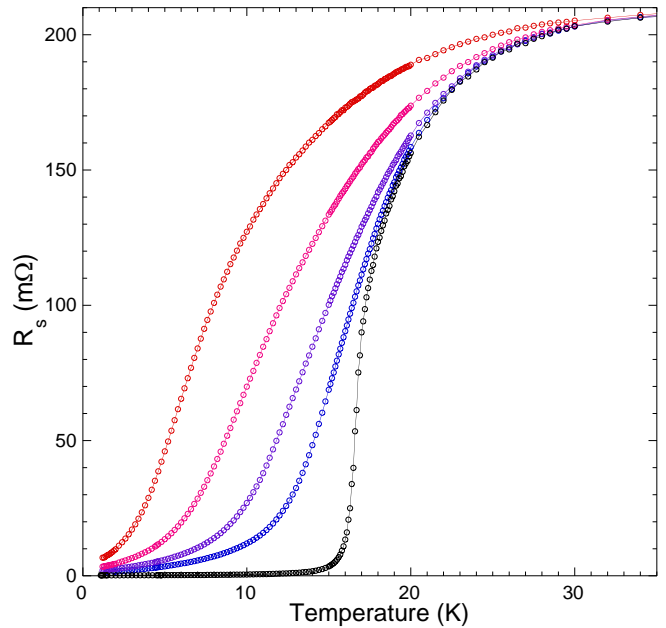}
\caption{\label{fig:InField} (Color online) $a$--$b$ plane surface resistance of a small ellipsoid of YBa$_2$Cu$_3$O$_{6.333}$ in an applied magnetic field.  The different traces show magnetic fields of 7, 3, 1, 0.25 and 0~T, from top to bottom.  At the lowest temperatures the uncertainty in surface resistance is $\pm 4 \mu\Omega$.  At higher temperatures it is $\pm$0.2\%.}
\end{figure}

The 8722ES is connected directly to the input coaxial line and provides input power levels up to -5~dBm.  We use several techniques to boost detection sensitivity, starting with the bandwidth-narrowing options built into the VNA.  Our 8722ES has Option~012, direct sampler access, allowing us to bypass the 16~dB insertion loss of the VNA's input directional coupler.  In addition, a peculiarity of the VNA's superheterodyne receiver allows substantial gains to be made by judicious preamplification.  The receiver's local oscillator is a high order ($n \approx 200$) harmonic comb, which brings in noise at all the mixer's many image frequencies.  Preamplification is carried out with a pair of Miteq AFS5 series amplifiers, each with $\approx25$~dB gain from 2 to 20~GHz and a noise figure (NF) $\approx 5$~dB.  These are followed by an HP8445B tunable YIG preselector, with insertion loss (IL) $\approx 5$~dB, that restricts the preamplifier gain to a 75~MHz band around the resonator frequency.   This greatly reduces the noise at the mixer's image frequencies.  Below 10~GHz, preamplification boosts the dynamic range of the VNA by close to the theoretical maximum of \mbox{(Gain - NF - IL)}~$\approx 40$~dB, tailing off at higher frequencies. At the low frequency end, the combination of direct sampler access and band-limited preamplification reduces the noise floor of the VNA to -165~dBm/Hz. This allows us to operate with low input power, or to make low-noise measurements on very lossy samples without having to increase coupling.

\section{Performance}

\begin{figure*}[t]
\includegraphics*[width=\textwidth]{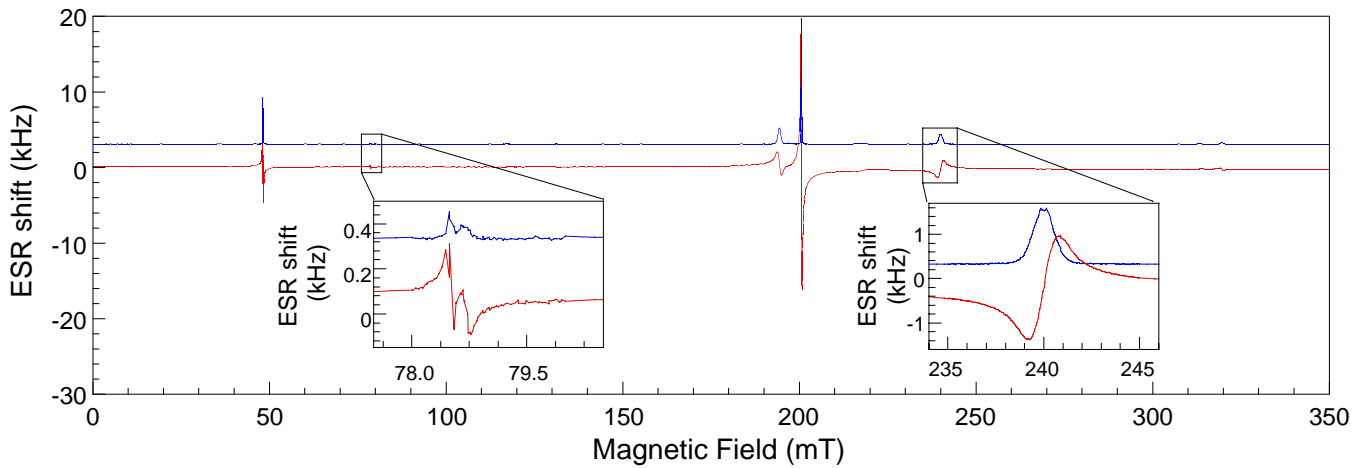}
\caption{\label{fig:ESR} (Color online)  Electron spin resonance (ESR) at 5.49~GHz in the smaller of the resonators.   ESR transitions are due to paramagnetic impurities in the rutile dielectric.  The upper trace is the resonator bandwidth $f_B$ and the lower trace is the frequency shift $\Delta f_0$ with respect to the resonator frequency in zero field.  Insets: Close-up views of small ESR transitions, demonstrating the signal-to-noise ratio of the measurement.  Bandwidth $f_B$ has been offset by a constant in each case to fit in the panels.  Frequency shift errors of 20~Hz correspond to resolution at the 4 parts~per~billion level.}
\end{figure*}

In this section we present representative data acquired with the surface impedance probe under a range of experimental conditions.  Fig.~\ref{fig:PbRxXsvsT} shows 5.49~GHz surface impedance data for a calibration sample, a small sphere of PbSn alloy.  By comparing the measured $f_B$ data to the d.c.\ resistivity of the PbSn alloy we can empirically determine the resonator constant $\Gamma$ to an accuracy of 2.5\%.  For a metal in the low frequency limit, we expect $R_s$ and $X_s$ to be equal in the normal state.  Indeed, this is often the best way to set the origin in $X_s$, as it cannot be measured directly.  However, the data in Fig.~\ref{fig:PbRxXsvsT} reveal a large discrepancy.  The disagreement is due to thermal expansion of PbSn, which contributes a significant error term $\Delta X_s^{\rm th}$ to the surface reactance.  Thermal expansion has the same effect on the reactance as a decrease in the penetration depth, so
\begin{equation}
\Delta X_s^{\rm th}(T) = - \omega \mu_0\Delta r(T) = -\omega \mu_0 r \frac{\Delta \ell(T)}{\ell}.
\end{equation}
For our spherical sample accurate dimensions were obtained by weighing with a microgram balance.  From the mass, 254~$\mu$g, we compute a radius  $r = 175.9$~$\mu$m.  Using thermal expansion data from the literature, \cite{thermalexpansion} a correction term $-\Delta X_s^{\rm th}(T)$ has been generated and added to the measured reactance signal.  The corrected $\Delta X_s$ data are shown as the solid line in Fig.~\ref{fig:PbRxXsvsT}.  We emphasize that there are no free parameters in this analysis.   The temperature dependence of the corrected $X_s$ data tracks $R_s(T)$ very accurately above 25~K, allowing the {\em absolute} reactance to be determined by matching $R_s$ and $X_s$ in that temperature range.    It should be pointed out that the thermal expansivity of Pb is one of the largest of all elements.  The thermal expansivities of oxides and transition metals such as Nb are typically over an order of magnitude smaller.\cite{thermalexpansion}  Nevertheless, it is always necessary to rule out thermal expansion as a potential source of error in a measurement of the surface reactance.  

Below 25~K, $R_s$ and $X_s$ diverge: this is a result of the increasing mean free path, which introduces relaxation and nonlocal effects into the surface impedance.  The superconducting surface impedance is enlarged in the inset of Fig.~\ref{fig:PbRxXsvsT}, showing the exponentially flat temperature dependence of $R_s$ and $X_s$ at low temperatures.

Next we turn to experiments in a static magnetic field.  Fig.~\ref{fig:InField} shows $a$--$b$ plane surface resistance at 5.49~GHz for a small ellipsoid of highly underdoped YBa$_2$Cu$_3$O$_{6.333}$, in magnetic fields of 0, 0.25, 1, 3 and 7~T.  These results are for illustrative purposes and will be published in detail elsewhere.  The high resolution data obtained from the new apparatus allow very precise determinations of the flux-flow resistivity and depinning frequency.  The ellipsoid measured here is 300~$\mu$m in diameter, but we are able to resolve the surface resistance at low temperature with an accuracy of $\pm 4~\mu\Omega$ at 5.49~GHz.  This is comparable to the resolution obtained in high frequency superconducting resonators, due to the high filling factor.

Finally, we present data from an electron spin resonance (ESR) experiment on the empty 5.49~GHz resonator, with the static magnetic field applied along the crystal $c$ direction.  $\Delta f_0$ and $\Delta f_B/2$ were measured using rapid frequency scans as the applied magnetic field was ramped between 0 and 350 mT, with the data plotted in Fig.~\ref{fig:ESR}.   As shown in the Appendix, the shifts in frequency and bandwidth are directly related to the complex magnetic susceptibility of the material.  One of the great advantages of cavity perturbation is the ability to measure both the real and imaginary parts of a complex response function simultaneously.
In this experiment, the ESR transitions are due to paramagnetic impurities in the rutile. They clearly illustrate the resolution of the technique and also serve as a caution, showing that paramagnetic impurities must be taken into account when designing the dielectric resonator --- significant loss of performance can occur if a zero-field ESR transition coincides with the resonant frequency of one of the operating modes.  Close inspection of the insets of Fig.~\ref{fig:ESR} reveals a frequency-shift uncertainty of 20~Hz, corresponding to frequency resolution at the 4 parts per billion level.  This level of noise is actually much higher than the sub-Hz frequency-shift errors we typically obtain with the apparatus, due to the fast frequency sweep rate used to acquire the data.
 
These experiments give a flavour for the utility and resolution of the rutile-resonator apparatus.  It is flexible enough to be put to many different uses, without modification, and is robust enough for reliable, everyday operation in the laboratory.

\section*{Acknowledgements}

We are most grateful to J.~C.~Gallop for bringing rutile resonators to our attention, to P.~Dosanjh for technical assistance, and to S.~Kamal for resistivity data on the PbSn alloy.  This research was funded by the National Science and Engineering Research Council of Canada and the Canadian Institute for Advanced Research.

\appendix

\section{Theory of Cavity Perturbation}
\label{app:CavPert}

In this appendix we show how cavity perturbation can be used  to obtain direct measurements of the surface
impedance $Z_s$, the permittivity $\epsilon$, and the permeability $\mu$ of small samples placed in the fields of a microwave resonator.\footnote{$Z_s$, $\epsilon$ and $\mu$ can be treated as scalars, because we usually align the symmetry axes of single crystal samples with the microwave field polarization.} The utility of the technique lies in its simplicity --- changes in the complex quantities $Z_s$, $\epsilon$ and $\mu$ are directly proportional to shifts in the resonant frequency $f_0$ and bandwidth $f_B$ of the microwave resonator.  

We use the methods of Altshuler\cite{altschuler63} and Ormeno~{\em et~al.}\cite{ormeno97} in the following.   The subscripts 1 and 2 denote the configurations before and after the perturbation, respectively.   Although a continuous wave source is used to sweep out the frequency response of the resonant modes, the experiment is effectively probing the free-decay response of the resonator. It is therefore convenient to use a  complex-exponential notation in which, for example, the time dependance of the magnetic field is given by
$\vec H(t)=\Re\{\textbf{H}e^{i\tilde{\omega}t}\}$.  Here \textbf{H} is a
complex phasor vector field and $\tilde{\omega}=\omega'+i\omega''$ is a
complex angular frequency with positive imaginary part.  The Fourier
transform of this time dependance is a Lorentzian centered at
$f_0=\omega'/2\pi$, with full width at half maximum
$f_B=\omega''/\pi$.

The integral over the cavity volume
\begin{equation}\label{eqn:VolInt}
  \int_V\big\{\textbf{E}_1\cdot\nabla\times\textbf{H}_2
    -\textbf{H}_2\cdot\nabla\times\textbf{E}_1
    +\textbf{H}_1\cdot\nabla\times\textbf{E}_2
    -\textbf{E}_2\cdot\nabla\times\textbf{H}_1\big\}\mathrm{d}V
\end{equation}
can be converted to a surface integral over the cavity walls using
the divergence theorem, resulting in 
\begin{equation}\label{eqn:SurfaceInt}
  \int_S[\textbf{H}_2\times\textbf{E}_1-\textbf{H}_1\times\textbf{E}_2]
    \cdot\hat{\textbf{n}}\mbox{\ }\mathrm{d}S,
\end{equation}
where $\hat{\textbf{n}}$ is the outward pointing normal to the
surface.  

The surface impedance relates the tangential
components of the electric and magnetic fields at the surface and is
formally defined by
\begin{equation}\label{eqn:ZsRel}
  \textbf{E}_t=Z_s\hat{\textbf{n}}\times\textbf{H},
\end{equation}
allowing \ref{eqn:SurfaceInt} to be rewritten as
\begin{equation}\label{eqn:SurfaceInt2}
  \int_S\big[Z_{s_1}\textbf{H}_2\times(\hat{\textbf{n}}\times\textbf{H}_1)
    -Z_{s_2}\textbf{H}_1\times(\hat{\textbf{n}}\times\textbf{H}_2)\big]
    \cdot\hat{\textbf{n}}\mbox{\ }\mathrm{d}S.
\end{equation}
Using the identity
$\textbf{A}\times(\textbf{B}\times\textbf{C})
  =\textbf{B}(\textbf{A}\cdot\textbf{C})-\textbf{C}(\textbf{A}\cdot\textbf{B})$
this becomes
\begin{eqnarray}
  \lefteqn{\int_S\Big\{Z_{s_1}\big[\hat{\textbf{n}}(\textbf{H}_2\cdot\textbf{H}_1)
    -\textbf{H}_1(\textbf{H}_2\cdot\hat{\textbf{n}})\big]}\hspace{0.4cm}\nonumber
  \\&&-Z_{s_2}\big[\hat{\textbf{n}}(\textbf{H}_1\cdot\textbf{H}_2)
    -\textbf{H}_2(\textbf{H}_1\cdot\hat{\textbf{n}}) \big]\Big\}
   \cdot\hat{\textbf{n}}\,\mathrm{d}S.\label{eqn:SurfaceInt3}
\end{eqnarray}
The high-frequency fields deep inside the metal surface vanish, which, by continuity of flux, 
requires $\hat{\textbf{n}}\cdot\textbf{H}_{1,2}=0$. We then obtain that
\begin{equation}\label{eqn:Zs}
  \int_S[\textbf{H}_2\times\textbf{E}_1-\textbf{H}_1\times\textbf{E}_2]
    \cdot\hat{\textbf{n}}\,\mathrm{d}S=-\int_S\Delta Z_s\textbf{H}_1\cdot\textbf{H}_2
    \mbox{\ }\mathrm{d}S,
\end{equation}
showing that \ref{eqn:VolInt} is directly related to a change in the surface impedance of the surface bounding the volume.

Our next step is to relate the surface impedance to the complex resonator frequency, and to changes in the electric and magnetic properties of the materials inside the resonator.   Faraday's law in phasor form is
\begin{equation}
  \nabla\times\textbf{E}_n = -i\tilde{\omega}_n
    \mu_n\textbf{H}_n,
    \label{eqn:Faraday}
\end{equation}
where $n = 1,2$.  Amp\`ere's law is usually written
\begin{equation}
\nabla\times\textbf{H}_n = \textbf{J}_n
    +i\tilde{\omega}_n\epsilon_n\textbf{E}_n,\label{eqn:Ampere1}
\end{equation}
but here we choose to absorb the conduction current density $\textbf{J}_n = \sigma_n \textbf{E}_n$ into a redefinition of the permittivity,\footnote{This reflects that fact that at finite frequency there is no fundamental difference between free and bound charge. To avoid spurious frequency shifts in the experiment, samples of highly conducting material should be positioned in regions of low electric field.}  $\epsilon_n \rightarrow \epsilon_n' = \epsilon_n - i \sigma_n/\omega$,   so that 
\begin{equation}
\nabla\times\textbf{H}_n =  i\tilde{\omega}_n\epsilon_n'\textbf{E}_n.\label{eqn:Ampere2}
\end{equation}

 Substituting
Eqs.~\ref{eqn:Faraday} and \ref{eqn:Ampere2} into \ref{eqn:VolInt} yields
\begin{eqnarray}
  \lefteqn{\int_V\big\{\textbf{E}_2\cdot\big(i\tilde{\omega}_1
    \epsilon'_1\textbf{E}_1\big)+\textbf{H}_1\cdot\big(i\tilde{\omega}_2\mu_2
    \textbf{H}_2\big)}\hspace{0cm}\nonumber
  \\&&-\textbf{H}_2\cdot\big(i\tilde{\omega}_1\mu_1\textbf{H}_1\big)-\textbf{E}_1
    \cdot\big(i\tilde{\omega}_2\epsilon'_2\textbf{E}_2\big)\big\}
    \mathrm{d}V\hspace{0.75cm}
  \\&=&-i\int_V\Big\{\big(\tilde{\omega}_2\epsilon'_2-\tilde{\omega}_1
    \epsilon'_1\big)\textbf{E}_1\cdot\textbf{E}_2\nonumber
  \\&&+\big(\tilde{\omega}_1\mu_1-\tilde{\omega}_2\mu_2\big)\textbf{H}_1\cdot
    \textbf{H}_2\Big\}\mathrm{d}V
  \\&=&i\tilde{\omega}_1\int_V\big[\Delta\epsilon'\textbf{E}_1\cdot\textbf{E}_2
  -\Delta\mu\textbf{H}_1\cdot\textbf{H}_2\big]\mathrm{d}V\nonumber
  \\&&+i\Delta\tilde{\omega}\int_V\big[\epsilon_2'\textbf{E}_1\cdot\textbf{E}_2
    -\mu_2\textbf{H}_1\cdot\textbf{H}_2\big]\mathrm{d}V.\label{eqn:VolIntFinal}
\end{eqnarray}

For  small perturbations
$\textbf{E}_1\approx\textbf{E}_2$ and
$\textbf{H}_1\approx\textbf{H}_2$. For a high $Q$ resonance
$\tilde{\omega}$ is predominantly real so that to good approximation the phase
difference  between $\textbf{E}$ and
$\textbf{H}$ is $\pm\frac{\pi}{2}$.  Without loss of
generality we set the phase of $\textbf{H}$ to zero, giving the following result:
\begin{eqnarray}
  \lefteqn{\int_V(\mu_2\textbf{H}_1\cdot\textbf{H}_2
    -\epsilon'_2\textbf{E}_1\cdot\textbf{E}_2)\mathrm{d}V}\nonumber
  \\&&\approx\int_V(\mu_2|H|^2+\epsilon'_2|E|^2)\mathrm{d}V=4U,\label{eqn:Energy}
\end{eqnarray}
where $U$ is the energy stored in the resonator.

Equating Eq.~\ref{eqn:VolIntFinal} to Eq.~\ref{eqn:SurfaceInt3}, substituting Eq.~\ref{eqn:Energy}, and
solving for $\Delta\tilde{\omega}$ yields the main cavity perturbation result:
\begin{eqnarray}
  \lefteqn{\Delta f_0 + i\Delta f_B/2\approx\bigg\{\frac{i}{2 \pi}\int_S\Delta Z_s\mathbf{H_1\cdot H_2}\mathrm{d}S}\nonumber
  \\&& - f_0\int_V[\Delta\mu
    \mathbf{H}_1\cdot\mathbf{H}_2+\Delta\epsilon'\mathbf{E}_1\cdot\mathbf{E}_2]
    \mathrm{d}V \bigg\}\bigg/4 U.\hspace{0.5cm}
  \label{eqn:cavitypert}
\end{eqnarray}


\end{document}